\documentclass[aip,rsi,reprint,graphicx]{revtex4-2}

\usepackage{graphicx}
\usepackage{dcolumn}
\usepackage{bm}
\usepackage[utf8]{inputenc}
\usepackage[T1]{fontenc}
\usepackage{mathptmx}
\usepackage{upgreek}
\usepackage{physics}
\usepackage{layouts}
\usepackage[colorlinks,plainpages=false,linkcolor=blue,urlcolor=blue,citecolor=blue,pdfpagemode=UseNone,pdfstartview=FitH]{hyperref}
\usepackage{lipsum}
\usepackage{relsize}
\usepackage{siunitx}
\usepackage{float}
\usepackage{caption}

\begin{document}
	
\title{The interplay between electron tunneling and Auger emission in a single quantum emitter weakly coupled to an electron reservoir}

\author{M. Zöllner}
\email{marcel.zoellner@uni-due.de}
\affiliation{Faculty of Physics and CENIDE, University of Duisburg-Essen, 47057 Duisburg, Germany}

\author{H. Mannel}
\affiliation{Faculty of Physics and CENIDE, University of Duisburg-Essen, 47057 Duisburg, Germany}
\author{F. Rimek}

\affiliation{Faculty of Physics and CENIDE, University of Duisburg-Essen, 47057 Duisburg, Germany}

\author{B. Maib}
\affiliation{Faculty of Physics and CENIDE, University of Duisburg-Essen, 47057 Duisburg, Germany}

\author{N. Schwarz}
\affiliation{Faculty of Physics and CENIDE, University of Duisburg-Essen, 47057 Duisburg, Germany}

\author{A. D. Wieck}
\affiliation{Chair of Applied Solid State Physics, Ruhr-Universität Bochum, 44780 Bochum, Germany}

\author{A. Ludwig}
\affiliation{Chair of Applied Solid State Physics, Ruhr-Universität Bochum, 44780 Bochum, Germany}

\author{A. Lorke}
\affiliation{Faculty of Physics and CENIDE, University of Duisburg-Essen, 47057 Duisburg, Germany}

\author{M. Geller}
\affiliation{Faculty of Physics and CENIDE, University of Duisburg-Essen, 47057 Duisburg, Germany}

\date{\today}


\begin{abstract}	
	In quantum dots (QDs) the Auger recombination is a non-radiative scattering process in which the optical transition energy of a charged exciton (trion) is transferred to an additional electron leaving the dot. Electron tunneling from a reservoir is the competing process that replenishes the QD with an electron again. Here, we study the dependence of the tunneling and Auger recombintaion rate on the applied electric field using high-resolution time-resolved resonance fluorescence (RF) measurements. With the given p-i-n diode structure and a tunnel barrier between the electron reservoir and the QD of $\SI{45}{\nano\meter}$, we measured a tunneling rate into the QD in the order of $\si{\per\milli\second}$. This rate shows a strong decrease by almost an order of magnitude for decreasing electric field, while the Auger emission rate decreases by a factor of five in the same voltage range. Furthermore, we study in detail the influence of the Auger recombination and the tunneling rate from the charge reservoir into the QD on the intensity and linewidth of the trion transition. Besides the well-known quenching of the trion transition, we observe in our time-resolved RF measurements a strong influence of the tunneling rate on the observed linewidth. The steady-state RF measurement yields a broadened trion transition of about $\SI{1.5}{\giga\hertz}$ for an Auger emission rate of the same order as the electron tunneling rate. In a non-equilibrium measurement, the Auger recombination can be suppressed, and a more than four times smaller linewidth of $\SI{340}{\mega\hertz}$ ($\SI{1.4}{\micro\electronvolt}$) is measured.
\end{abstract}

\maketitle

\section{Introduction}
A promising stationary quantum bit (qubit) is the spin of an electron (or hole) in a solid state environment\cite{Kimble2008, Atatuere2006}, where the quantum state of the spin\cite{Benjamin2009} can be transferred by a spin-photon interface\cite{Press2008, Yilmaz2010, Atatuere2018, Javadi2018} to a photon that serves as a flying qubit. The connection between both qubits can be realised by the charged exciton state (the trion)\cite{Delteil2015, Bernien2013} in a single self-assembled quantum dot (QD).\cite{Petroff2001, Bimberg1997} Therefore, long coherence times and highly indistinguishable photons\cite{Matthiesen2012, Atatuere2018, Lio2022} are needed. Previous findings showed spin and charge noise as the main dephasing mechanisms\cite{Houel2012, Kuhlmann2013}, which lead to a broadening of the natural linewidth of the exciton and trion transition. The influence of other possible mechanisms on the linewidth and coherence of the single photons, such as the non-radiative Auger effect \cite{Park2014, Kurzmann2016a, Lochner2020}, the radiative Auger effect \cite{Loebl2020}, or the internal photoeffect \cite{Lochner2021}, are still under investigation. \\
We study here in time-resolved resonance fluorescence the influence of the electron tunneling and the non-radiative Auger recombination on the applied electric field and measure simultaneously linewidth and intensity of the trion transition. The QD is weakly coupled to an electron reservoir\cite{Luyken1999} with tunneling rates in the order of $\si{\per\milli\second}$. This rate shows a strong decrease for increasing electric field, while the Auger scattering rate decrease by a factor of five in the same voltage range. The tunneling rate $\gamma_\mathrm{in}$ and the electron emission rate $\gamma_\mathrm{e}$ by the Auger recombination can be tuned by the laser intensity to the same order of magnitude to investigate the interplay between the electron tunneling and the Auger emission on linewidth and intensity of the trion transition. In this regime of competing rates, where an electron is emitted from the dot (by Auger) and an electron is recharged from the reservoir (by tunneling), we measure in a steady-state resonance fluorescence measurement an artificially broadened linewidth of $\SI{1.5}{\giga\hertz}$ and a reduced trion intensity. In a non-equilibrium transient RF, where the Auger recombination can be suppressed, we obtain a four times smaller value for the linewidth of $\SI{340}{\mega\hertz}$ ($\SI{1.4}{\micro\electronvolt}$). The resulting dephasing time $T_{2}$ of $\SI{957}{\pico\second}$ is in good agreement with previously observed values for self-assembled QDs.\cite{Kurzmann2016a} These results demonstrate the strong influence of the Auger recombination on the optical properties of the charged exciton transition, which may help to improve the fabrication of optimized single photon emitters as well as spin-to-charge and spin-to-photon conversion devices.

\section{Sample Structure and Measurement Setup}
The measurements were performed on a single self-assembled (InGa)As QD, grown in a Stranski–Krastanov process\cite{Baskaran2012} by molecular beam epitaxy. The layer of QDs is embedded in a p-i-n diode structure with an electron reservoir consisting of a highly n-doped GaAs:Si layer, a $\SI{45}{\nano\meter}$ thick (AlGa)As tunneling barrier, and a highly p-doped GaAs layer as the epitaxial gate\cite{Ludwig2017}, see the supplementary material for more details. An indium-flush\cite{Wasilewski1999} during the growth process limits the height of the QDs so that their exciton emission wavelength is between $\SI{900}{\nano\meter}$ and $\SI{1000}{\nano\meter}$. A voltage, applied between the electron reservoir and the epitaxial gate allows us to charge the QD with single electrons from the reservoir.\cite{Geller2019} Furthermore, we can use the quantum confined Stark effect\cite{Li2000} to tune the QD states in resonance with our excitation laser. To read out the charge states, we use resonance fluorescence spectroscopy in a confocal microscope setup at a sample temperature of $\SI{4.2}{\kelvin}$. To distinguish the QD photons from the laser photons, i.e. to suppress the backscattered laser light, we use the cross-polarization method with a maximum suppression of $10^{7}$ laser photons.\cite{Yilmaz2010, Kurzmann2016a}

\section{Results and Discussion}
\begin{figure}[H]
	\centering
	\includegraphics{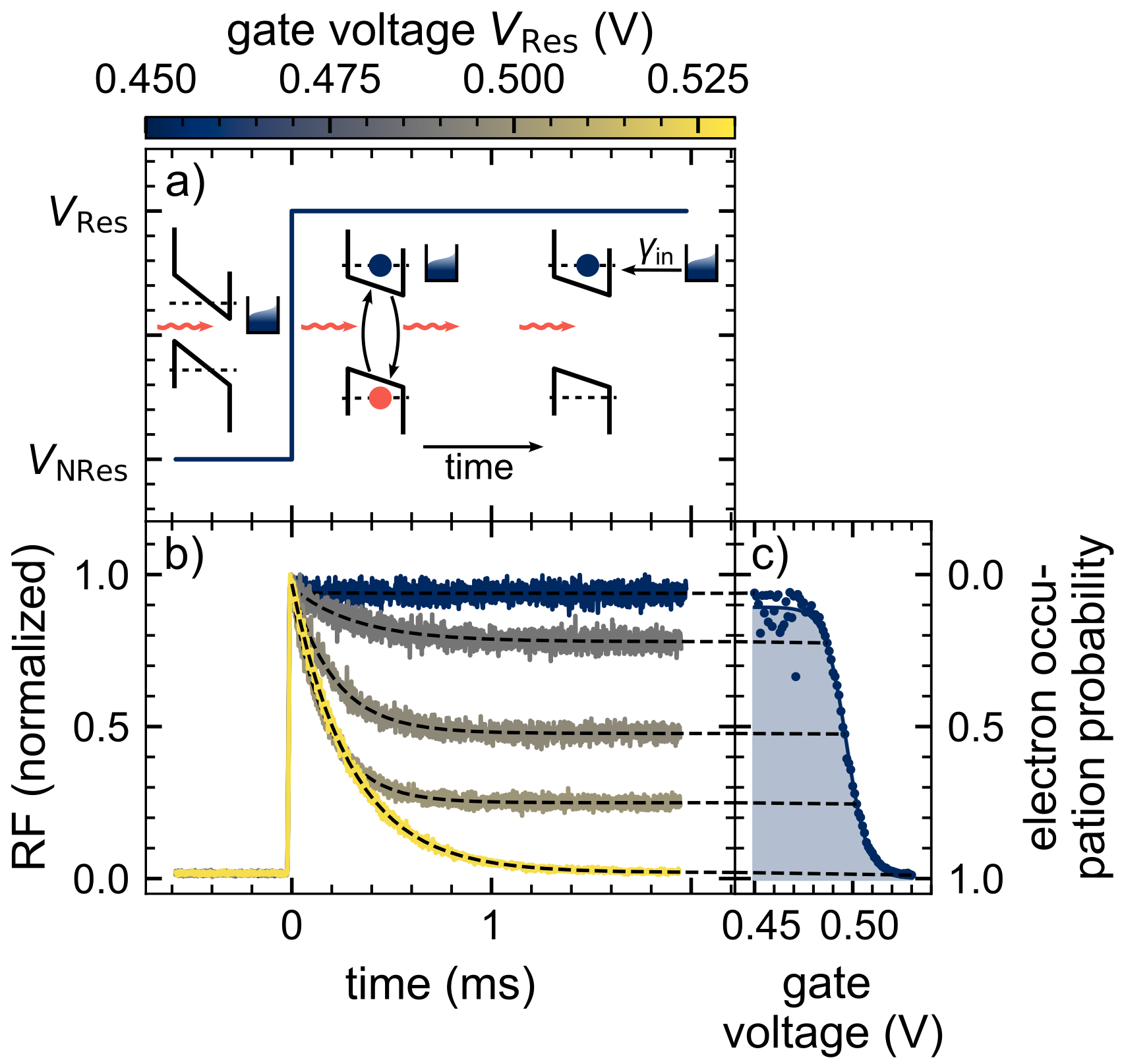}
	\caption{a) Time-resolved N-shot measurement scheme with a $\SI{2}{\milli\second}$ pulse of the gate voltage. The pulses from $V_\mathrm{NRes}$ to $V_\mathrm{Res}$ set the electron reservoir out of and into resonance with the $s_{1}$-ground state of the QD. b) Exciton RF intensity for a $\SI{1}{\mu\second}$ binning time during the pulse. An exponential decrease of the intensity is observed, caused by an electron tunneling from the electron reservoir into the QD. c) Occupation probability of a single electron in the QD. The shape corresponds to the Fermi distribution of the electron reservoir, with a temperature of $\SI{4.2}{\kelvin}$.}
	\label{fig.:measurement_scheme}
\end{figure}

We will show in the following the gate voltage dependent tunneling dynamics for a single electron tunneling event. We use a time-resolved gate voltage N-shot pulse scheme with a pulse duration of $\SI{2}{\milli\second}$, as shown in Fig.~\ref{fig.:measurement_scheme}a). The continuous-wave laser with an excitation intensity of $\SI[per-mode=symbol]{2e-2}{\micro\watt\per\square\micro\meter}$ will not be pulsed. The pulses from $V_\mathrm{NRes}$ to $V_\mathrm{Res}$ set the electron reservoir out of and into resonance with the s-shell ground state of the dot (see small insets in Fig.~\ref{fig.:measurement_scheme}a)) to tunnel an electron into and out of the QD. The non-resonant gate voltage $V_\mathrm{NRes}$ is set to zero voltage. 

The resonant gate voltage $V_\mathrm{Res}$ is tuned between $\SI{0.45}{\volt}$ and $\SI{0.7}{\volt}$. The laser frequency must be shifted with the applied electric field from $\SI{327.376}{\tera\hertz}$ to $\SI{327.382}{\tera\hertz}$ to account for the quantum confined Stark effect. The N-shot pulse scheme is applied approximately $50.000$ times per data point and the RF intensity as a function of time is determined by collecting the detected photons with a bin width of $\SI{1}{\mu\second}$. While at time $t = 0$, the dot is always empty (preparation of an empty QD by $V_\mathrm{NRes} = \SI{0}{\volt}$), so that the exciton emission is at a maximum. For gate voltages $V_\mathrm{Res}$ above $\SI{0.497}{\volt}$, the energy of the one-electron state shifts through the Fermi edge, increasing the probability of an electron tunneling into the dot and blocking the neutral exciton transition due to the singly charged QD.\cite{Luyken1999, Kurzmann2016b} This reduces the exciton RF intensity exponentially over time in our N-shot averaged experiment, shown in Fig.~\ref{fig.:measurement_scheme}b), where the color of the transient represents the gate voltage $V_\mathrm{Res}$. The exponential decrease of the exciton intensity $I(t)$ can be described by the following relation (shown in Fig.~\ref{fig.:measurement_scheme}b) with black dashed lines):

\begin{equation}
	I(t) = I_{0} \left(\frac{\gamma_\mathrm{in}}{\gamma_\mathrm{m}} \cdot e^{-\gamma_\mathrm{m}t} + \frac{\gamma_\mathrm{out}}{\gamma_\mathrm{m}}\right) \text{ ,}
	\label{eq.:tunneling_fit_function}
\end{equation}

which is derived from a simple two state rate equation.\cite{Kurzmann2016b} Here, the relaxation rate $\gamma_\mathrm{m}$ is given by the electron tunneling rate into the QD $\gamma_\mathrm{in}$ and the electron tunneling rate out of the QD $\gamma_\mathrm{out}$:

\begin{equation}
	\gamma_\mathrm{m} = \gamma_\mathrm{in} + \gamma_\mathrm{out} \text{ .}
	\label{eq.:exciton_relaxation_rate}
\end{equation}

\begin{figure}[H]
	\centering
	\includegraphics{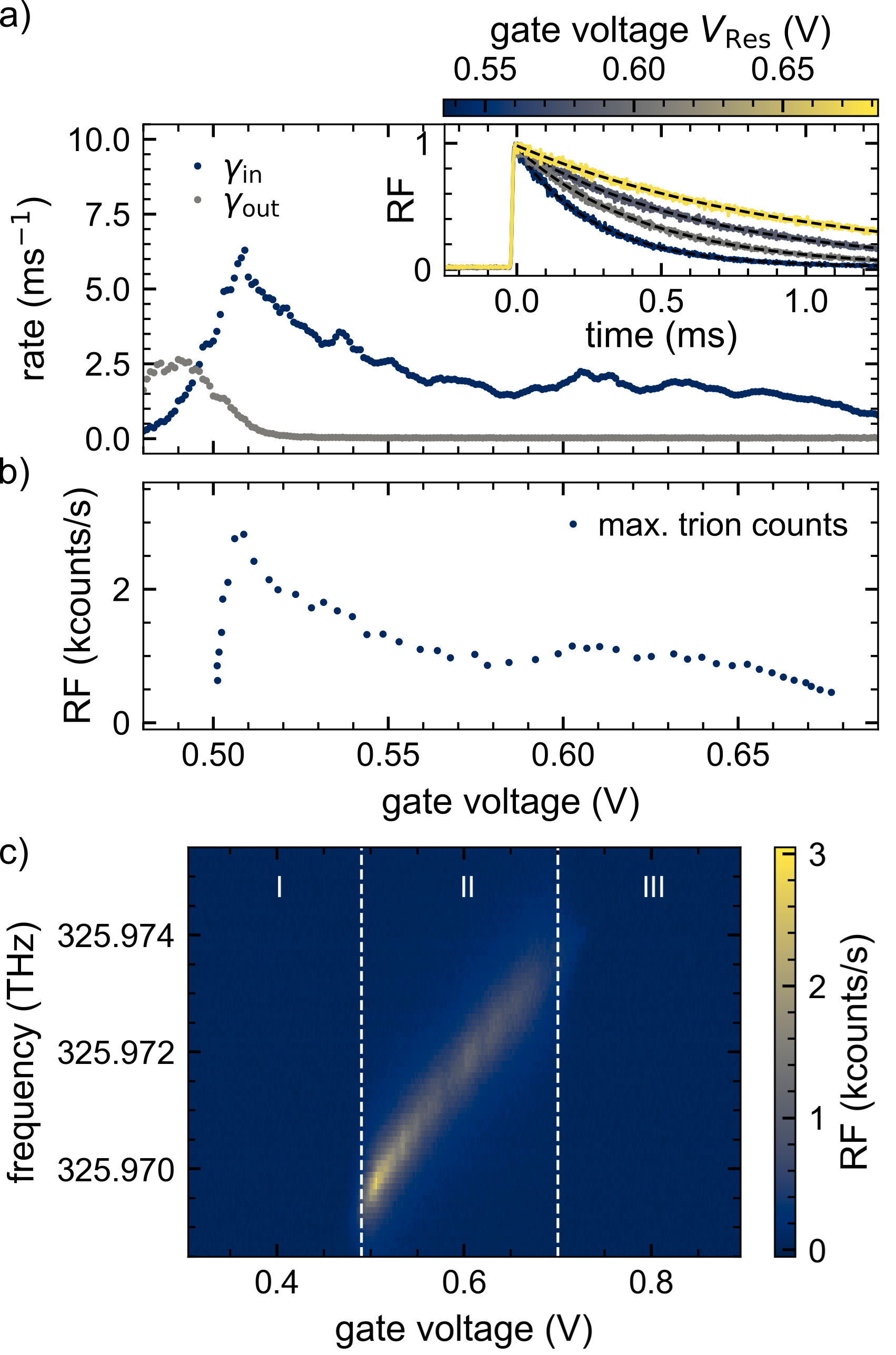}
	\caption{a) Single electron tunneling rates into ($\gamma_\mathrm{in}$, blue) and out of ($\gamma_\mathrm{out}$, grey) the QD as a function of the gate voltage $V_\mathrm{Res}$. The inset shows four exponentially decaying transients, recorded at the color-coded gate voltages. The tunneling rates were calculated from all transients using Eq.~\eqref{eq.:tunneling_fit_function}. b) Maximum trion counts per second extracted from steady-state RF measurements (shown in Fig.~\ref{fig.:gatevoltage_dependence_olaf}c)) at $\SI[per-mode=symbol]{1.1e-4}{\micro\watt\per\square\micro\meter}$ (which corresponds to a trion occupation probability of $n_{X^-}(0) = 0.012$) as a function of the gate voltage. c) Color-coded trion RF intensity of the QD as a function of gate voltage and excitation frequency. The quantum confined Stark effect shifts the resonance frequency linearly with the gate voltage.}
	\label{fig.:gatevoltage_dependence_olaf}
\end{figure}

The long-term limit of this function ($t\rightarrow\infty$) gives us the steady-state occupation probability as a function of the gate voltage, shown in Fig.~\ref{fig.:measurement_scheme}c). The blue line describes the Fermi function of the electron reservoir fitted to the data. From this we can obtain the temperature of the QD sample's electron reservoir of $\SI{4.2}{\kelvin}$, which is in excellent agreement with the liquid helium temperature. \\
Figure \ref{fig.:gatevoltage_dependence_olaf}a) shows the tunneling rates $\gamma_\mathrm{in}$ and $\gamma_\mathrm{out}$, as obtained from Eq.~\eqref{eq.:tunneling_fit_function} and shown exemplarily in Fig.~\ref{fig.:measurement_scheme}b) and in the upper-right inset in Fig.~\ref{fig.:gatevoltage_dependence_olaf}a), as a function of the resonant gate voltage $V_\mathrm{Res}$. For low gate voltages, we observe that the tunneling rate $\gamma_\mathrm{in}$ increases with increasing gate voltage up to a gate voltage $V_\mathrm{Res} = \SI{0.51}{\volt}$. The spin degeneracy of the empty QD gives a factor of two for the maximum tunneling rate into the QD ($\SI{6.3}{\per\milli\second}$ at $\SI{0.51}{\volt}$) in comparison to the maximum tunneling rate out of the QD ($\SI{2.6}{\per\milli\second}$ at $\SI{0.49}{\volt}$); as shown previously in Kurzmann et al.\cite{Kurzmann2016b, Beckel2014}. However, above the maximum tunneling rate we observe an unusual strong decrease of the tunneling rate into the dot by almost an order of magnitude ($\SI{6.3}{\per\milli\second}\rightarrow\SI{0.7}{\per\milli\second}$) in the gate voltage range from $\SI{0.51}{\volt}$ to $\SI{0.70}{\volt}$, while the tunneling rate out of the QD remains constant zero as expected, due to the Pauli exclusion principle blocking carriers to tunnel into the occupied electron reservoir. In addition, we observe resonance-like features in the tunneling rate $\gamma_\mathrm{in}$, which are most likely due to local defects in the vicinity of the QD.\cite{Efros2016, Nguyen2013, Kerski2021} This would allow resonant electron tunneling through the barriers, as observed before by Könemann et al.\cite{Koenemann2002}. The overall trend of a decreasing tunneling rate $\gamma_\mathrm{in}$ for increasing gate voltage $V_\mathrm{Res}$ between $\SI{0.51}{\volt}$ and $\SI{0.70}{\volt}$ reflects a high-resolution measurement of the density of states in the electron reservoir, as the tunneling rate is given by:\cite{Kurzmann2016b}

\begin{equation}
	\gamma_\mathrm{in} = d_\mathrm{in} \cdot \Gamma \cdot f(E) \text{ ,}
	\label{eq.:definition_tunneling_rate}
\end{equation}

with the degeneracy of the final state $d_\mathrm{in}$ and the Fermi distribution $f(E)$. As discussed by Beckel et al.\cite{Beckel2014}, the electron transition rate through the tunnel barrier $\Gamma$ contains, according to Fermi's golden rule\cite{Fermi1974, Dirac1927}, the density of states of the initial state (electron reservoir) and the final state (QD). Since the spin degeneracy of the final states for tunneling into the QD is energy independent with $d_\mathrm{in} = 2$ and the Fermi distribution is approximately one at gate voltages above $\SI{0.51}{\volt}$, the energy dependence of the electron tunneling can therefore only come from the transition rate through the tunnel barrier $\Gamma(E)$.

With this strong dependence of the tunneling rate $\gamma_\mathrm{in}$ in mind, we will show how this rate has a strong influence on the trion intensity and linewidth in an steady-state gate voltage dependent measurement. \\
Figure \ref{fig.:gatevoltage_dependence_olaf}c), displays the color-coded RF intensity of the trion recombination as a function of the excitation frequency and the gate voltage. The three areas represent the empty (I), the singly charged (II) and the doubly charged QD (III). These measurements were performed as steady-state measurement, so that they represent the trion intensity in the long-term limit ($t\rightarrow\infty$). In addition to the linear quantum confined Stark effect of the trion transition, we can also observe the gate voltage dependence of the maximum trion intensity, depicted in Fig.~\ref{fig.:gatevoltage_dependence_olaf}b). These measurements show a decreasing trion intensity as a function of the gate voltage, almost identical to the shape of the tunneling rate into the QD, shown in Fig.~\ref{fig.:gatevoltage_dependence_olaf}a). The resonance-like features, previously discussed for the tunneling rate $\gamma_\mathrm{in}$, are also observed in the maximum trion intensity. \\
Furthermore, we observe a decrease of the maximum trion intensity by almost a factor of nine ($\SI{2.5}{\kilo counts\per\second}\rightarrow\SI{0.3}{\kilo counts\per\second}$), when increasing the gate voltage from $\SI{0.51}{\volt}$ to $\SI{0.69}{\volt}$. This is also in very good agreement with the decrease of the tunneling rate $\gamma_\mathrm{in}$. Here, it should be taken into consideration that the trion steady-state intensity (Eq.~\eqref{eq.:auger_fit_function} for $t\rightarrow\infty$) is given by the tunneling rates, and also by the electron emission through the non-radiative Auger effect:\cite{Kurzmann2016a}

\begin{equation}
	I(t\rightarrow\infty) =I_{0} \left(\frac{\gamma_\mathrm{in}}{\gamma_\mathrm{in}+\gamma_\mathrm{out}+\gamma_\mathrm{e}}\right) \text{ ,}
	\label{eq.:definition_damping_probability}
\end{equation}

with the Auger emission rate $\gamma_\mathrm{e}$. During the non-radiative Auger effect, the energy released in the recombination of the electron-hole pair is transferred to the second electron, causing it to leave the QD. Only when another electron has tunneled from the electron reservoir into the QD, the trion transition can be optically driven again. In self-assembled QDs, this effect was first shown by Kurzmann et al.\cite{Kurzmann2016a} and later explored in more detail by Lochner et al.\cite{Lochner2020} and Mannel et al.\cite{Mannel2021}. The developed time-resolved N-shot pulse scheme allows us to determine the Auger emission rate $\gamma_\mathrm{e}$ as well as the tunneling rates very accurately. In this case we pulse the resonant trion excitation laser with an acousto-optic modulator (AOM)\cite{Gordon1966} and a pulse duration of $\SI{2}{\milli\second}$. We again tune the gate voltage between $\SI{0.45}{\volt}$ and $\SI{0.7}{\volt}$. At the start of each cycle, the laser is turned off. Then the chosen range of gate voltages assures that the QD is tuned off and the quantum dot is occupied with a single electron. By this preparation with an electron in the dot, the undisturbed trion transition can immediately excited after the laser is switched on. Within the timescale of the electron emission by the Auger recombination, however, a decreasing transient arises due to the non-radiative Auger effect, which ejects the electron from the QD and thus partially quenches the trion resonance. This transient follows a similar time dependence as discussed in Eq.~\eqref{eq.:tunneling_fit_function}, when the Auger emission rate $\gamma_\mathrm{e}$ is taken into account in addition to the tunneling rates:\cite{Kurzmann2016a}

\begin{equation}
	I(t) = I_{0} \cdot \left(\frac{\gamma_\mathrm{e}}{\gamma_\mathrm{m}} \cdot e^{-\gamma_\mathrm{m}t} + \frac{\gamma_\mathrm{in}+\gamma_\mathrm{out}}{\gamma_\mathrm{m}}\right) \text{ ,} 
	\label{eq.:auger_fit_function}
\end{equation}

with the trion relaxation rate: 

\begin{equation}
	\gamma_\mathrm{m} = \gamma_\mathrm{in} + \gamma_\mathrm{out} + \gamma_\mathrm{e} \text{ ,} 
	\label{eq.:trion_relaxation_rate}
\end{equation}

\begin{figure}[H]
	\centering
	\includegraphics{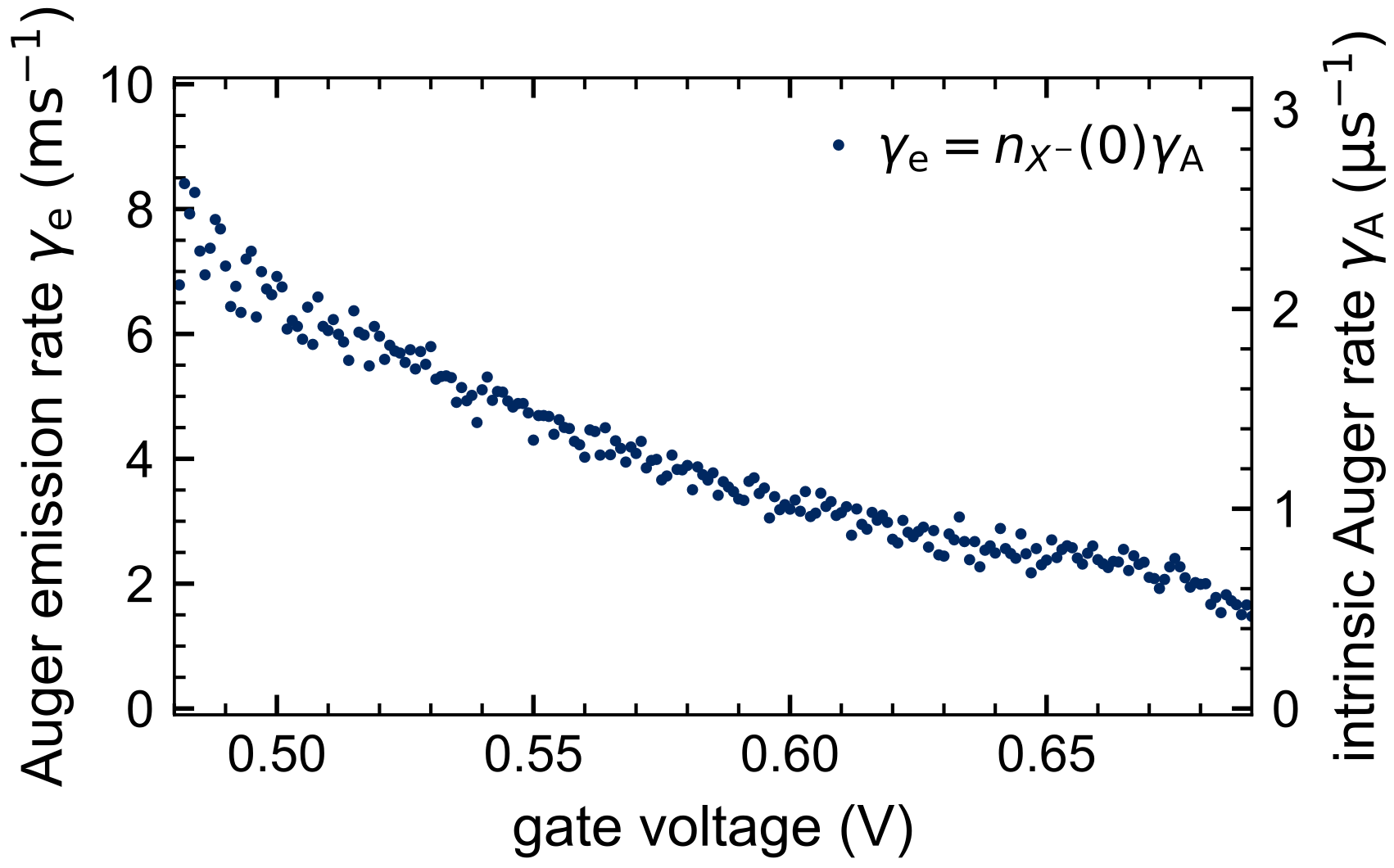}
	\caption{Auger emission rate $\gamma_\mathrm{e}$ as a function of the gate voltage, derived from Eq.~\eqref{eq.:auger_fit_function}, using the tunneling rates $\gamma_\mathrm{in}$ and $\gamma_\mathrm{out}$, and the Auger emission rate $\gamma_\mathrm{e}$ as free fit parameters. For this time-resolved N-shot measurement the laser was pulsed with a trion excitation intensity of $\SI[per-mode=symbol]{2.8e-5}{\micro\watt\per\square\micro\meter}$ (which corresponds to a trion occupation probability of $n_{X^-}(0) = 0.003$).}
	\label{fig.:gatevoltage_auger}
\end{figure}

The Auger emission rate for a fixed trion excitation intensity of $\SI[per-mode=symbol]{2.8e-5}{\micro\watt\per\square\micro\meter}$ (which corresponds to a trion occupation probability of $n_{X^-}(0) = 0.003$) as a function of the gate voltage is shown in Fig.~\ref{fig.:gatevoltage_auger}. We observe that the Auger emission rate $\gamma_\mathrm{e}$ decreases by a factor of five ($\SI{8.4}{\per\milli\second}\rightarrow\SI{1.7}{\per\milli\second}$) with increasing gate voltage. In comparison to the tunneling rate the behavior of the Auger emission rate is rather smooth and has no resonances or rapid slope changes. These are not expected either, since the Auger effect is not affected by the electron transition through the tunnel barrier $\Gamma$ or by defects in the environment of the QD. The tunneling rates $\gamma_\mathrm{in}$ and $\gamma_\mathrm{out}$, on the other hand, which result from the same fit to the exponentially decaying trion transients, show the same behavior as the tunneling rates that were determined from the exciton data in Fig.~\ref{fig.:gatevoltage_dependence_olaf}a) (see the supplementary material).

\begin{figure}[H]
	\centering
	\includegraphics{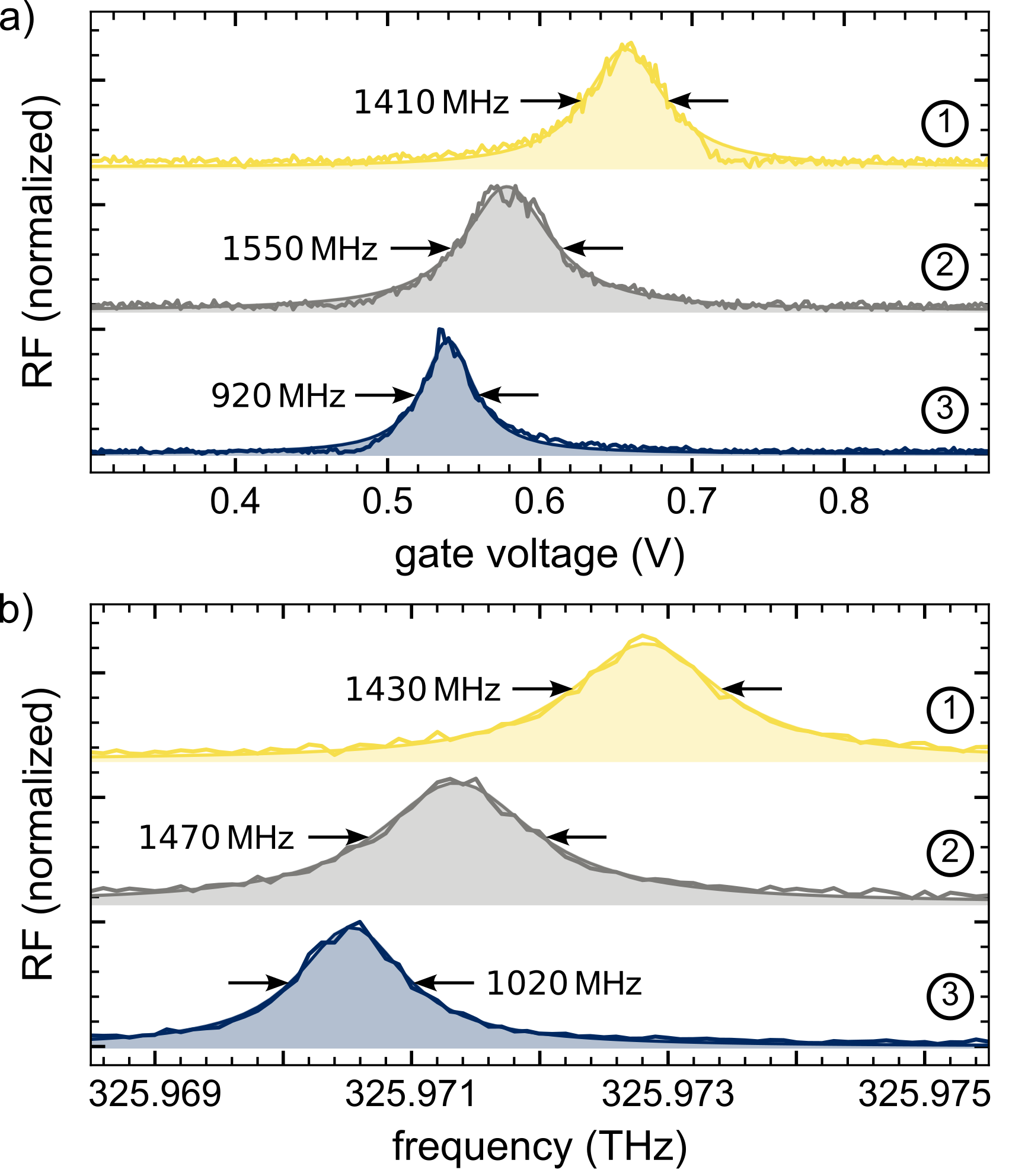}
	\caption{a) Three exemplary RF resonances with different excitation frequencies \textcircled{\scriptsize1} $\SI{325.9706}{\tera\hertz}$, \textcircled{\scriptsize2} $\SI{325.9714}{\tera\hertz}$ and \textcircled{\scriptsize3} $\SI{325.973}{\tera\hertz}$ extracted from the full scan shown in Fig.~\ref{fig.:gatevoltage_dependence_olaf}c), with normalized RF trion intensity as a function of the gate voltage (horizontal cuts). The area under the curve represents a Lorentz fit to the data. b) Three exemplary RF resonances with different gate voltages \textcircled{\scriptsize1} $\SI{0.54}{\volt}$, \textcircled{\scriptsize2} $\SI{0.578}{\volt}$ and \textcircled{\scriptsize3} $\SI{0.656}{\volt}$ extracted from the full scan shown in Fig.~\ref{fig.:gatevoltage_dependence_olaf}c), with normalized RF trion intensity as a function of the excitation frequency (vertical cuts). The colored area under the curve represents a Lorentz fit to the data.}
	\label{fig.:linewidth_dependence_olaf}
\end{figure}

In Fig.~\ref{fig.:linewidth_dependence_olaf}a) the linewidth of the trion recombination as a function of the gate voltage are exemplarily shown, for three laser excitation frequencies (horizontal cuts through Fig.~\ref{fig.:gatevoltage_dependence_olaf}c)) (\textcircled{\scriptsize1} $\SI{325.9706}{\tera\hertz}$, \textcircled{\scriptsize2} $\SI{325.9714}{\tera\hertz}$ and \textcircled{\scriptsize3} $\SI{325.9730}{\tera\hertz}$). Since these measurements were performed in steady-state, the resonances are broadened due to mechanisms, which empty the QD non-radiatively, such as the Auger effect\cite{Kurzmann2016a} or the internal photoeffect\cite{Lochner2021}. It can be observed that with increasing excitation frequency the linewidth increases between $\SI{920}{\mega\hertz}$ at a tunneling rate $\gamma_\mathrm{in}$ of $\SI{3}{\per\milli\second}$ and an Auger emission rate $\gamma_\mathrm{e}$ of $\SI{5.1}{\per\milli\second}$ and $\SI{1410}{\mega\hertz}$ at a tunneling rate $\gamma_\mathrm{in}$ of $\SI{1.7}{\per\milli\second}$ and an Auger emission rate $\gamma_\mathrm{e}$ of $\SI{2.4}{\per\milli\second}$. In the supplementary material we show that the linewidth of the trion emission also follows the electron tunneling rate $\gamma_\mathrm{in}$ very well. Since for these measurements the frequency as well as the excitation intensity is consistent, here the broadening of the trion resonance is given by the gate voltage dependent ratio between the tunneling rates $\gamma_\mathrm{in}(V_\mathrm{g})$, $\gamma_\mathrm{out}(V_\mathrm{g})$ and the Auger emission rate $\gamma_\mathrm{e}(V_\mathrm{g})$, according to Eq.~\eqref{eq.:definition_damping_probability}. A measurement of the trion linewidth in a gate voltage scan for nearly equal electron tunneling and Auger emission rates must therefore be treated with caution. \\
However, the same caution has to taken even if the gate voltage is fixed and the linewidth should be determined by a frequency scan as vertical cuts through Fig.~\ref{fig.:gatevoltage_dependence_olaf}c). Three of such vertical cuts are show in Fig.~\ref{fig.:linewidth_dependence_olaf}b). 

Now the excitation frequency at a set of fixed gate voltages, respectively fixed tunneling rate $\gamma_\mathrm{in}$, $\gamma_\mathrm{out}$, is varied from $\SI{325.9685}{\tera\hertz}$ up to $\SI{325.9755}{\tera\hertz}$. Due to the fixed gate voltage, the broadening mechanism by the changing tunneling rates, described above, is ruled out. However, it should be noted that the occupation probability $n(\Delta\omega)$ depends on the excitation frequency:\cite{Muller2007}

\begin{equation}
	n(\Delta\omega) = \frac{1}{2}\frac{\Omega_{0}^{2}T_{1}/T_{2}}{\Delta\omega + 1/T_{2}^{2} + \Omega_{0}^2T_{1}/T_{2}} \text{ ,} 
	\label{eq.:definition_occupation_probability}
\end{equation}

with the detuning $\Delta\omega = \omega_{0}-\omega = 2\pi(\nu_{0} - \nu)$, between the excitation frequency $\nu$ and the resonance frequency $\nu_{0}$. The excitation power is expressed in terms of the Rabi frequency $\Omega_{0}$. The times $T_{1}$ and $T_{2}$ are the lifetime of the excited trion state and the dephasing time, respectively. \\
Therefore, in Fig.~\ref{fig.:linewidth_dependence_olaf}b) a broadening of the trion resonance is still observed, now due to the Auger emission rate $\gamma_\mathrm{e}(\Delta\omega) = n(\Delta\omega)\gamma_\mathrm{A}$ that depends on the average trion occupation probability, cf.~Eq.~\eqref{eq.:definition_damping_probability}. To determine the intrinsic trion linewidth, we use the trion intensity in a pulsed measurement scheme at the beginning of the pulse ($t = \SI{0}{\milli\second}$); shown in detail in the supplementary material. The resulting linewidth $\Delta\nu$ of $\SI{340}{\mega\hertz}$ is about a factor of four narrower than the linewidth of the steady-state trion resonance and corresponds to a dephasing time $T_{2}$ of $\SI{957}{\pico\second}$. \\
Using Eq.~\eqref{eq.:definition_damping_probability} and (\ref{eq.:definition_occupation_probability}), an excitation frequency of $\SI{325.9714}{\tera\hertz}$ (curve \textcircled{\scriptsize2} in Fig.~\ref{fig.:linewidth_dependence_olaf}b)) and the previously measured values ($\gamma_\mathrm{in}=\SI{1.7}{\per\milli\second}$, $\gamma_\mathrm{out}=\SI{0.03}{\per\milli\second}$, $\gamma_\mathrm{A}=\SI{1.2}{\per\micro\second}$ and $T_{2}=\SI{957}{\pico\second}$), an occupation in resonance of $n_{X^-}(0) = 0.012$ is obtained, which is in perfect agreement with the used laser intensity of $\SI[per-mode=symbol]{1.1e-4}{\micro\watt\per\square\micro\meter}$ in Fig.~\ref{fig.:gatevoltage_dependence_olaf}b-c) and Fig.~\ref{fig.:linewidth_dependence_olaf}.

\section{Conclusion}
In summary, for a single self-assembled (InGa)As QD, coupled to an electron reservoir by a rather thick tunnel barrier of $\SI{45}{\nano\meter}$ thickness, we observed a strong dependence of the tunneling rate into the QD, of the order of $\si{\per\milli\second}$, on the applied electric field. The tunneling rate decreases by almost an order of magnitude for increasing gate voltage, while the Auger emission rate decrease by a factor of five in the same voltage range. The varying tunneling rate, as well as the Auger effect, affects the trion transition and its amplitude as well as the linewidth in steady-state measurements significantly. In the regime of equal rate for the electron emission by the Auger recombination and the electron tunneling into the dot, we determined in a steady-state resonance fluorescence measurement an artificially broadened linewidth and a reduced trion intensity. In a non-equilibrium RF transient, where the Auger recombination can be suppressed, we obtain a four times smaller value for the linewidth of $\SI{340}{\mega\hertz}$ ($\SI{1.4}{\micro\electronvolt}$), which is in good agreement with previous results on a different self-assembled QD.\cite{Kurzmann2016a} This shows that the linewidth of the trion resonance measured in steady-state should always be interpreted with caution. However, for much larger tunneling rates into the QD in relation to the Auger emission rate, the Auger effect can be neglected (cf.~Eq.~\eqref{eq.:definition_damping_probability}) and the trion transition should not be artificially broadened.

 \begin{acknowledgments}
	This work was funded by the Deutsche Forschungsgemeinschaft (DFG, German Research Foundation) – Project-ID 278162697 – SFB 1242, and the individual research grant No. 383065199 and by the Mercator Research Center Ruhr (MERCUR) of Stiftung Mercator. A. Lu. and A. D. W. acknowledges support by DFG-TRR160, BMBF - QR.X KIS6QK4001, and the DFH/UFA CDFA-05-06. \\
\end{acknowledgments}

The data that support the findings of this study are available from the corresponding author upon reasonable request.

\bibliography{tunneling_dynamics}

\end{document}


\maketitle

\newpage

\section{Band stucture of the sample}
\begin{figure}[H]
	\centering
	\includegraphics{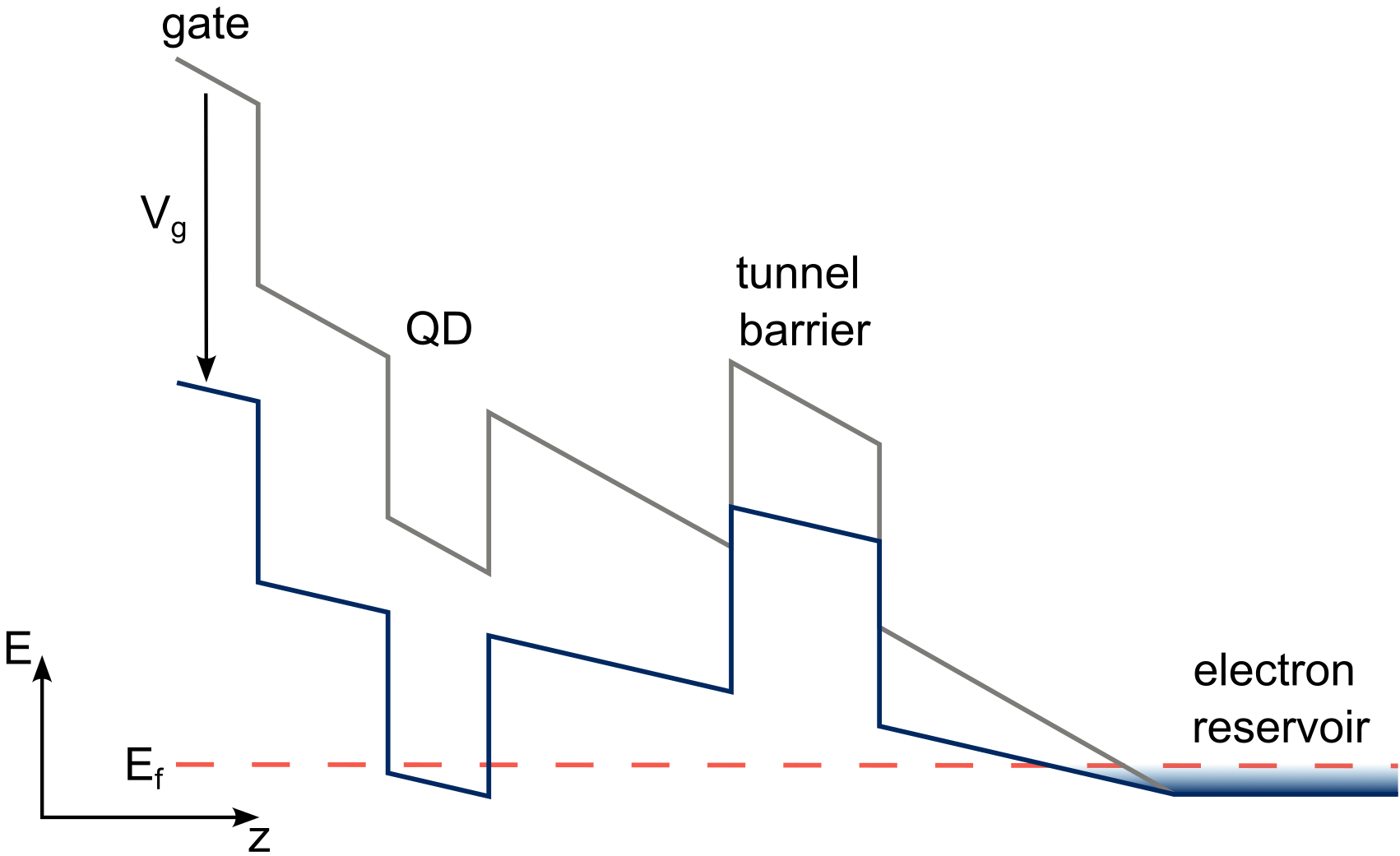}
	\caption{Schematic conduction band structure (grey: $V_\mathrm{g}=0$, blue: $V_\mathrm{g}>0$) as a function of the growth direction z, with the Fermi energy of the electron reservoir $E_\mathrm{f}$. With an applied electric field ($V_\mathrm{g}$) the conduction band is tilted in respect to the electron reservoir with a lever arm of $15.2$, so that the energy states of the QD can be loaded with electrons from the reservoir. (Inspired by Lochner et al.\cite{Lochner2019})}
	\label{fig.:sample_structure}
\end{figure}

\noindent In order to charge the QD states, the Fermi energy of the electron reservoir must be energetically higher then the QD state, so that a tunneling probability for an electron is given. In the diode structure used here, this is realized by applying a voltage between the gate contact and the electron reservoir, shown in Fig.~\ref{fig.:sample_structure}. Therefore, the conduction band is tilted with respect to the reservoir with a lever arm of $15.2$. 

\noindent In Fig.~2c), the gate voltage for different excitation frequencies is tuned. Since the optical trion transition is only allowed if an electron has already occupied the $s_{1}$-ground state of the QD, this resonant measurement can be used to determine the voltages at which an electron can tunnel through the tunnel barrier into the QD. In addition, it is possible to switch between the tunneling probabilities by pulsing the gate voltage, which can be used to determine the tunneling rates (shown in Fig.~2a)). \\
In the transients, which are generated by the time-resolved pulsed N-shot measurements at the trion transition, not only the Auger emission rate is included in the exponential decay, but also the tunneling rates, as shown in the fit function Eq.~(5). Therefore, this measurement can be used to confirm the gate voltage dependent tunneling rates measured by pulsed exciton resonance fluorescence, shown in Fig.~2a). For this purpose, $\gamma_\mathrm{in}$ and $\gamma_\mathrm{out}$ are chosen as a free fit parameter and adjusted together with the Auger emission rate $\gamma_\mathrm{e}$ to the respective transients.

\begin{figure}[H]
	\centering
	\includegraphics{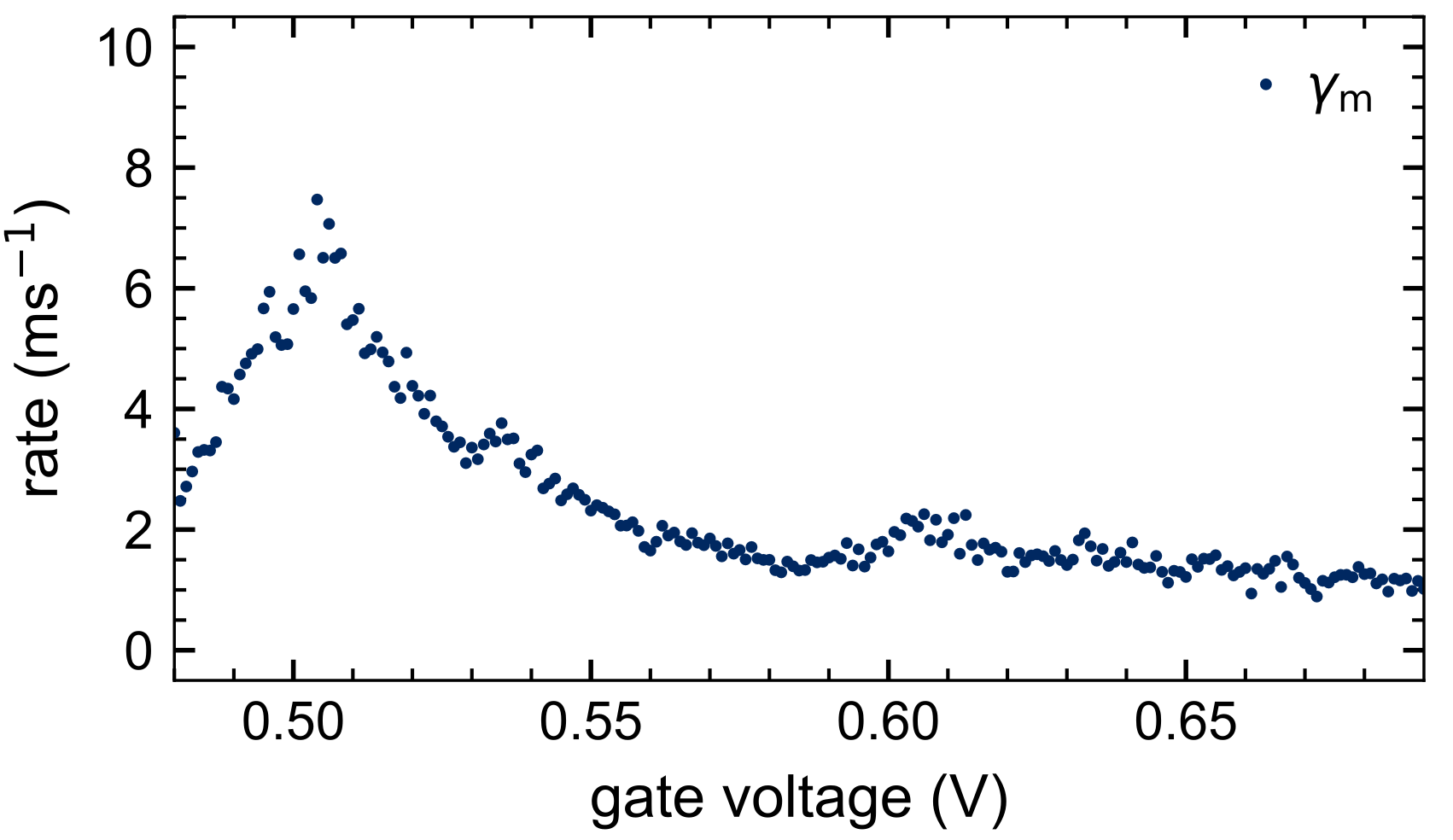}
	\caption{Electron tunneling rate $\gamma_\mathrm{m} = \gamma_\mathrm{in} + \gamma_\mathrm{out}$ as a function of the gate voltage $V_\mathrm{Res}$. The tunneling was measured indirect via the optical trion transition and the Auger emission rate $\gamma_\mathrm{e}$.}
	\label{fig.:tunneling_dependence_auger_olaf}
\end{figure}

\noindent As shown in Fig.~\ref{fig.:tunneling_dependence_auger_olaf}, the result are identical to the rate measured directly via the exciton transition (Fig.~2a)). It should be noted that in this measurement the rates $\gamma_\mathrm{in}$ and $\gamma_\mathrm{out}$ cannot be determined independently, because the function is overdefined.

\section{Occupation probability of the excited trion state}
\begin{figure}[H]
	\centering
	\includegraphics{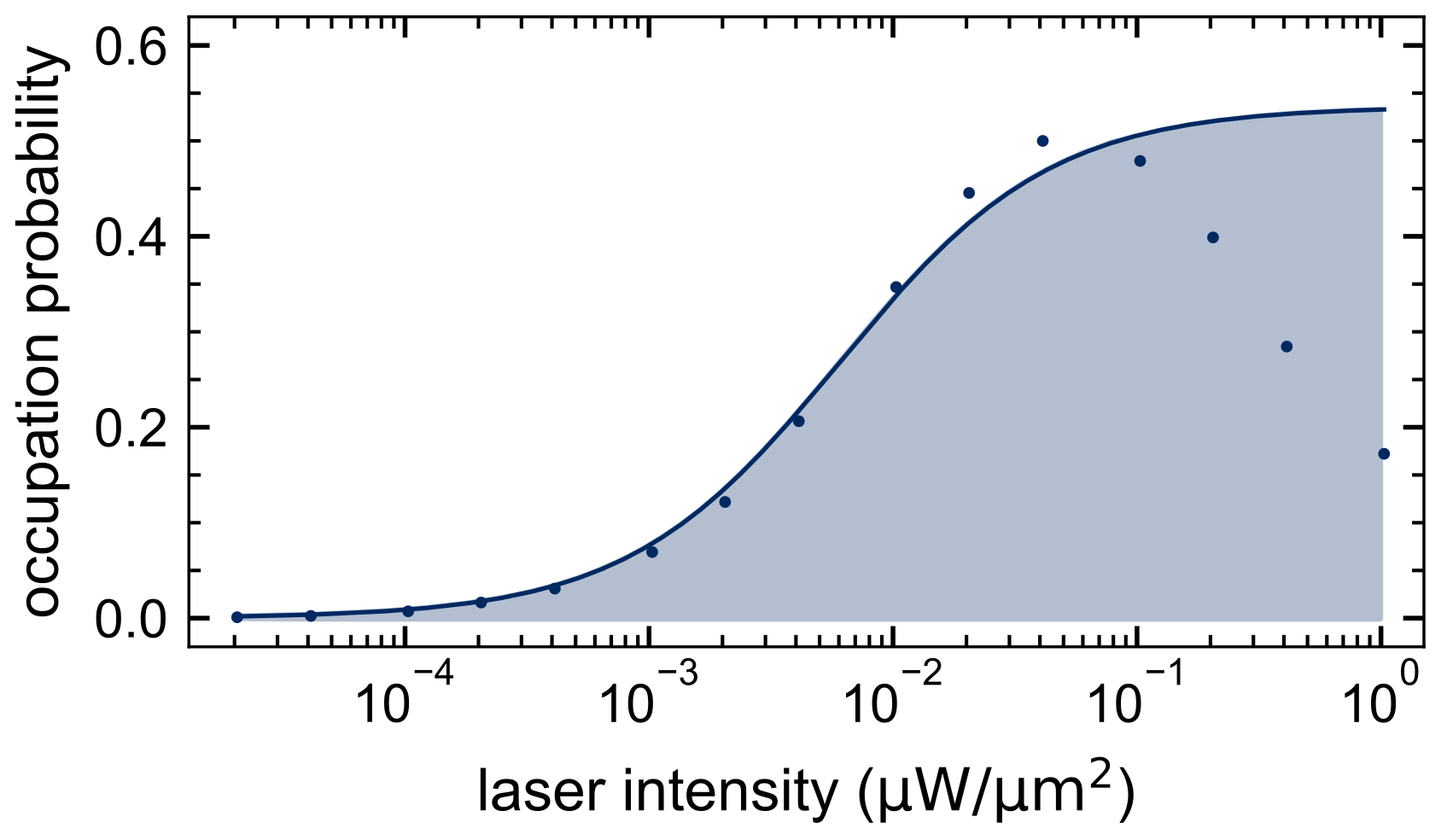}
	\caption{Laser intensity dependent occupation probability of the excited trion state. The data were acquired using the time-resolved pulsed measurement scheme. The occupation probability is given by the normalized counts in the trion transient at time $t=\SI{0}{\milli\second}$. The blue line indicates the fit to the data according to Eq.~\eqref{eq.:trion_saturation}.}
	\label{fig.:trion_saturation}
\end{figure}

\noindent Using time-resolved pulsed measurements of the trion transition, as shown by Kurzmann et al.\cite{Kurzmann2016a}, the excitation laser intensity dependent initial RF intensity ($t=\SI{0}{\milli\second}$) can be determined. Fig.~\ref{fig.:trion_saturation} shows the occupation probability of the excited trion state, which results from the normalized initial intensity, as a function of the excitation laser intensity. The fit indicated by the blue line has the form:

\begin{equation}
	I(x) = I_{0}\left(\frac{\alpha x}{\beta + \alpha x}\right) \text{ ,}
	\label{eq.:trion_saturation}
\end{equation}

\noindent which is a simplified form of Eq.~(7) for $\Delta\omega = 0$, and with $x = \Omega_{0}^{2}$, $\alpha = T_{1}/T_{2}$ and $\beta = 1/T_{2}^{2}$. 

\section{Gate-voltage dependent Full Width at Half Maximum (FWHM)}
\noindent In Fig.~2c), the color-coded RF intensity shows the relationship between the excitation frequency and the gate voltage for the trion transition. In order to convert the gate voltage into a frequency, a linear function is fitted to the Stark shift. The resulting Stark shift for $\SI{1}{\giga\hertz}$ is about $\SI{44.45}{\milli\volt}$.

\begin{figure}[H]
	\centering
	\includegraphics{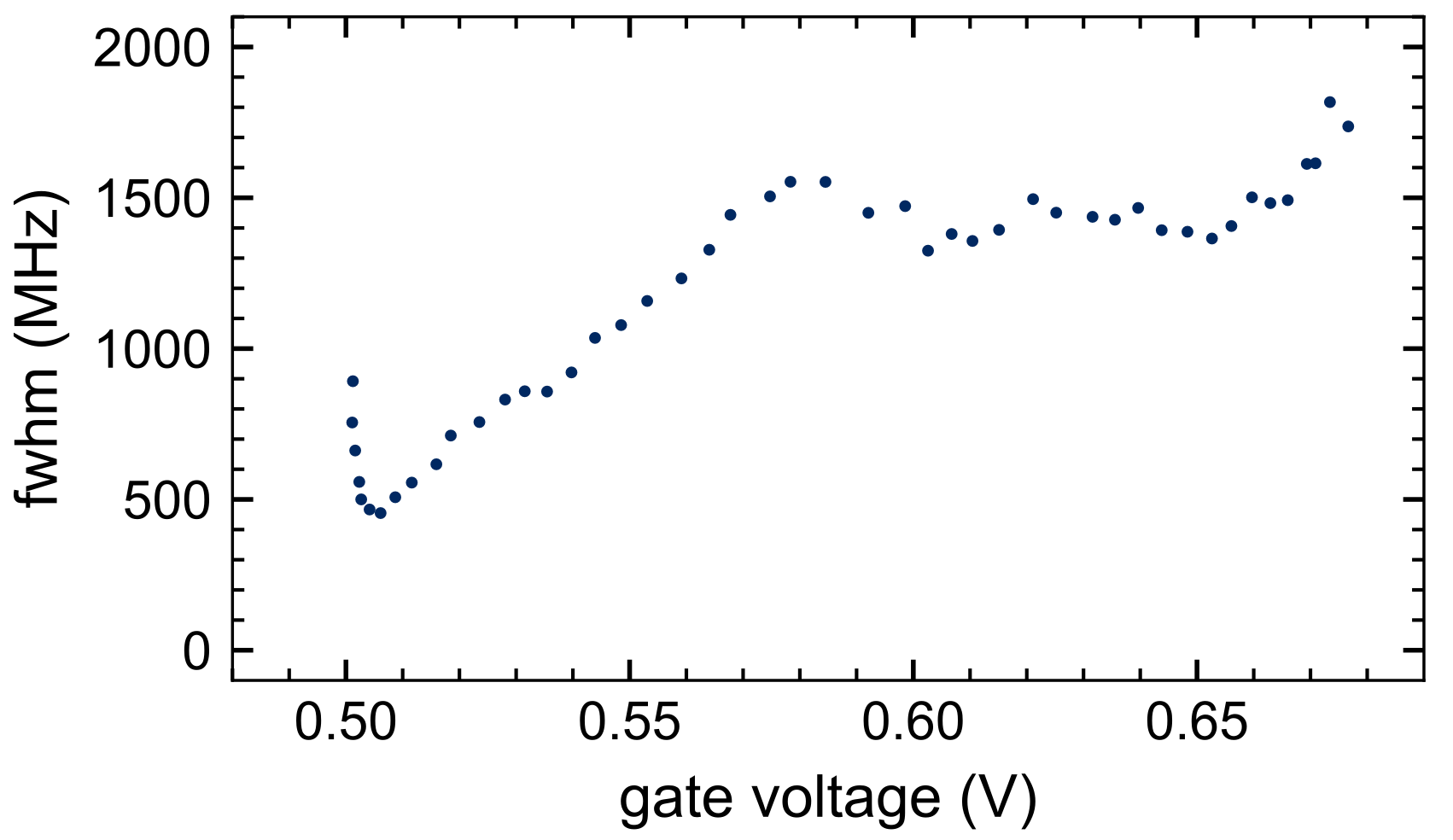}
	\caption{Full width at half maximum (fwhm) as a function of the gate voltage, calculated with a Lorentz fit to the steady-state measurements shown in Fig.~2c) and Fig.~4. The gate voltage dependent fwhm closely follows the inverse of the gate voltage dependent tunneling rate into the QD.}
	\label{fig.:gatevoltage_fwhm}
\end{figure}

\noindent For each gate voltage scan, we can calculate with a Lorentz fit the maximum trion counts Fig.~2c) and the Full Width at Half Maximum Fig.~\ref{fig.:gatevoltage_fwhm}. Since the linewidth of the trion transition depends on the ratio between the tunneling rate into the QD and the Auger emission rate as described in Eq.~(4), we observe that the shape of the FWHM as a function of gate voltage is inverse related to the tunneling rate $\gamma_\mathrm{in}$.

\section{Narrow trion linewidth without the Auger recombination}
\begin{figure}[H]
	\centering
	\includegraphics{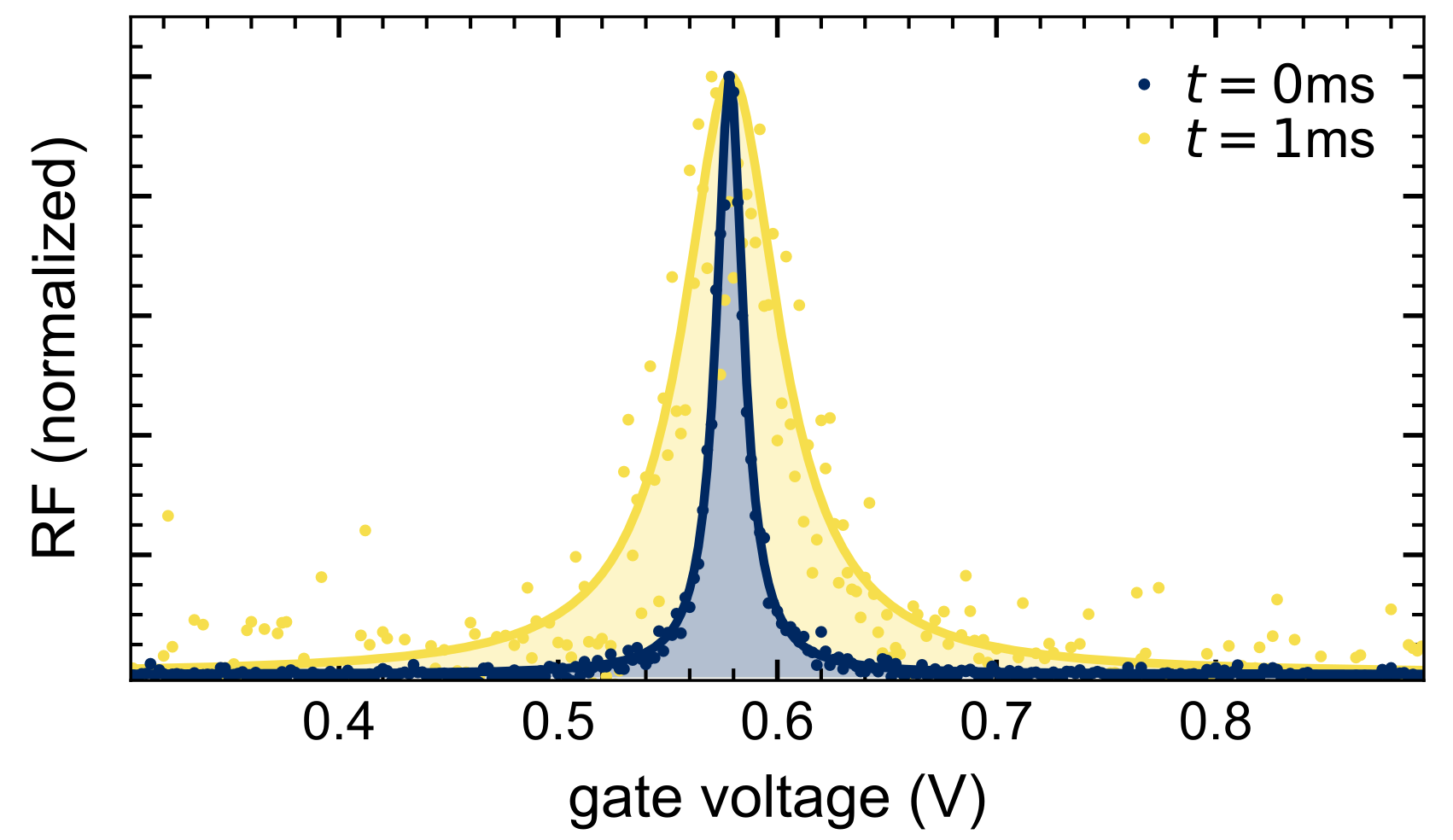}
	\caption{Time-resolved initial trion intensity (normalized) as a function of gate voltage for an excitation frequency of $\SI{325.9714}{\tera\hertz}$, extracted from non-equilibrium measurements. For $t=\SI{0}{\milli\second}$, processes that broaden the resonance in Fig.~4 such as Auger emission, are excluded. With the measured linewidth of $\SI{340}{\mega\hertz}$ (blue line), a dephasing time $T_{2}$ of $\SI{957}{\pico\second}$ results (for $\Omega_{0}\rightarrow0$, $\Delta\nu = 2 / 2\pi T_{2}$ applies). For $t=\SI{1}{\milli\second}$ (yellow line) the line is already broadend ($\SI{1190}{\mega\hertz}$) by a factor of three.}
	\label{fig.:gatevoltage_auger_i0}
\end{figure}

\noindent In order to determine the intrinsic trion linewidth without the Auger recombination, time-resolved pulsed-laser RF measurements on the trion transition were performed. These contain not only the Auger emission rate $\gamma_\mathrm{e}$, but also the linewidth. To obtain the undisturbed linewidth of the trion transition, the initial RF trion intensity $I_{0}$ has to be plotted versus the gate voltage. Figure \ref{fig.:gatevoltage_auger_i0} shows the undisturbed RF intensity and the RF intensity after $t = \SI{1}{\milli\second}$ for an excitation frequency of $\SI{325.9714}{\tera\hertz}$. 

\noindent We observed that the resonances in non-equilibrium measurement are significantly narrower with a linewidth of $\SI{340}{\mega\hertz}$. This shows that the ratio between the probability for tunneling into the QD and the Auger recombination, broaden the steady-state trion RF resonance up to a factor of four.

\section{Simulation of the trion resonance}
\noindent Equation (4) and (7) can be used to simulate the resonance of the trion transition in steady-state (which is broadened by the Auger effect) and in a non-equilibrium situation in a pulsed measurement at $t=0$, where the trion resonance is not broadened yet. 

\begin{figure}[H]
	\centering
	\includegraphics{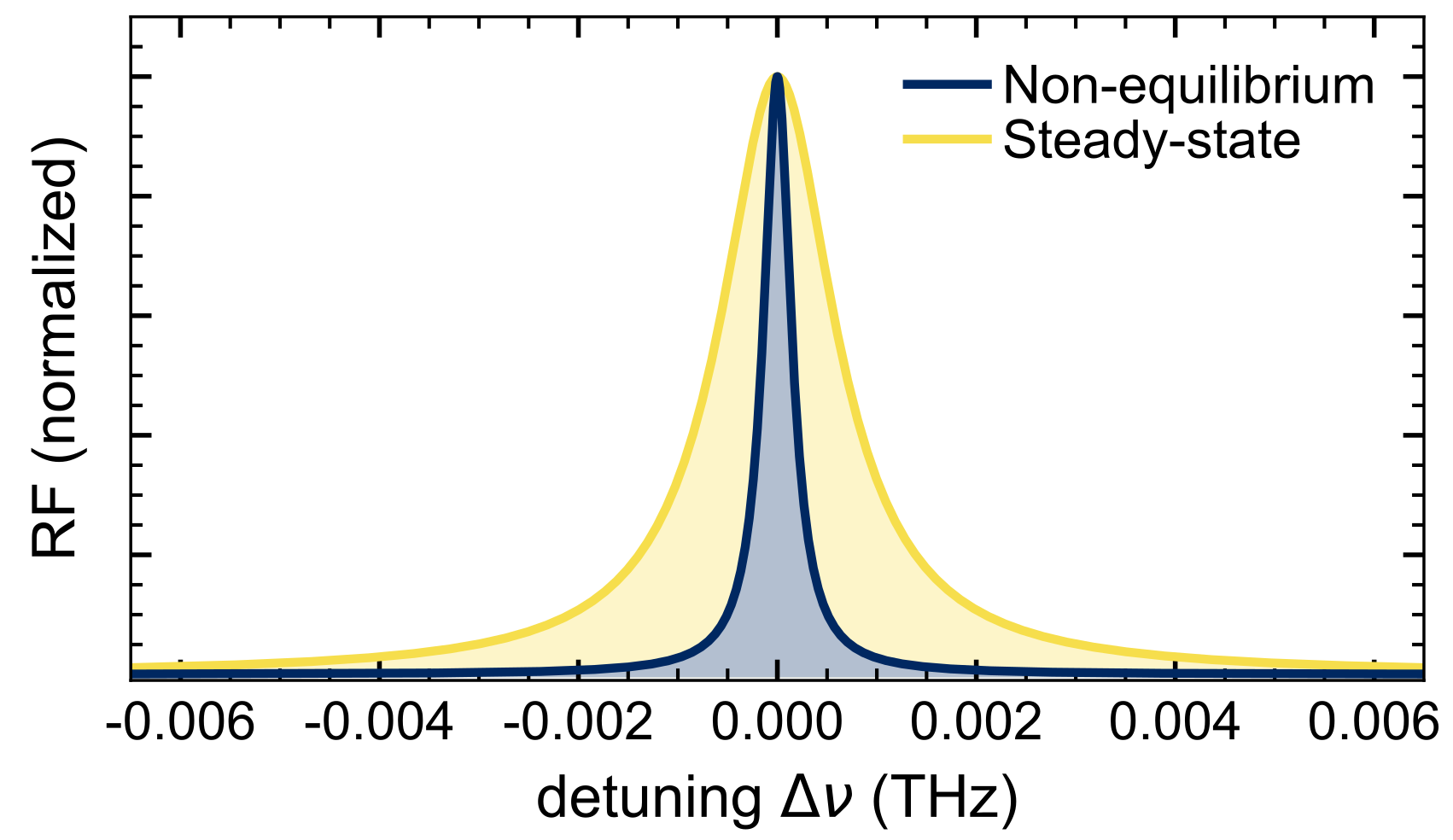}
	\caption{Calculation of the trion intensity distribution for a steady-state measurement (shown in Fig.~4b)) and a non-equilibrium measurement (shown in Fig.~\ref{fig.:gatevoltage_auger_i0}). For the calculation the tunneling rates $\gamma_\mathrm{in}$, $\gamma_\mathrm{out}$, the intrinsic Auger rate $\gamma_\mathrm{A}$, the trion occupation $n(\Delta\omega)$ and the dephasing time $T_{2}$ were taking to account. The unbroadened line shows a linewidth of $\SI{340}{\mega\hertz}$ and the broadend line due to the Auger effect shows a linewidth of $\SI{1390}{\mega\hertz}$, which is in good agreement with the measurement ($\SI{1470}{\mega\hertz}$).}
	\label{fig.:theory_intensity}
\end{figure}

\noindent The two simulated trion resonances are shown in Fig.~\ref{fig.:theory_intensity} and, together with the values measured in the main text for the tunneling rates $\gamma_\mathrm{in}$, $\gamma_\mathrm{out}$, the intrinsic Auger rate $\gamma_\mathrm{A}$, the trion occupation $n(\Delta\omega)$ and the dephasing time $T_{2}$, give almost identical linewidths as in the measurements.

\bibliography{tunneling_dynamics}